\documentclass[twocolumn,prb,superscriptaddress,showpacs,floatfix]{revtex4}

\usepackage{graphicx}
\usepackage{amssymb}
\usepackage{amsmath}
\usepackage{amsfonts}

\begin{document}

\title{Optical probing of spin dynamics of two-dimensional and bulk electrons\\
        in a GaAs/AlGaAs heterojunction system}

\author{P.~J.~Rizo}
\affiliation{Zernike Institute for Advanced Materials, University of Groningen,\\
Nijenborgh 4, NL-9747AG Groningen, The Netherlands}
\author{A.~Pugzlys}
\altaffiliation[Current address: ]{Photonics Institute,
Vienna University of Technology, Gusshausstrasse 27/387,
1040 Vienna, Austria}
\affiliation{Zernike Institute for Advanced Materials, University of Groningen,\\
Nijenborgh 4, NL-9747AG Groningen, The Netherlands}
\author{A.~Slachter}
\author{S.~Z.~Denega}
\affiliation{Zernike Institute for Advanced Materials, University of Groningen,\\
Nijenborgh 4, NL-9747AG Groningen, The Netherlands}
\author{D.~Reuter}
\affiliation{Angewandte Festk\"{o}rperphysik,
Ruhr-Universit\"{a}t Bochum, D-44780 Bochum, Germany}
\author{A.~D.~Wieck}
\affiliation{Angewandte Festk\"{o}rperphysik,
Ruhr-Universit\"{a}t Bochum, D-44780 Bochum, Germany}
\author{P.~H.~M.~van~Loosdrecht}
\affiliation{Zernike Institute for Advanced Materials, University of Groningen,\\
Nijenborgh 4, NL-9747AG Groningen, The Netherlands}
\author{C.~H.~van~der~Wal}
\affiliation{Zernike Institute for Advanced Materials, University of Groningen,\\
Nijenborgh 4, NL-9747AG Groningen, The Netherlands}

\date{5 Oct. 2009}

\begin{abstract}
We present time-resolved Kerr rotation measurements of
electron spin dynamics in a GaAs/AlGaAs heterojunction
system that contains a high-mobility two-dimensional
electron gas (2DEG). Due to the complex layer structure of
this material the Kerr rotation signals contain information
from electron spins in three different layers: the 2DEG
layer, a GaAs epilayer in the heterostructure, and the
underlying GaAs substrate. The 2DEG electrons can be
observed at low pump intensities, using that they have a
less negative g-factor than electrons in bulk GaAs regions.
At high pump intensities, the Kerr signals from the GaAs
epilayer and the substrate can be distinguished when using
a barrier between the two layers that blocks intermixing of
the two electron populations. This allows for stronger
pumping of the epilayer, which results in a shift of the
effective g-factor. Thus, three populations can be
distinguished using differences in g-factor. We support
this interpretation by studying how the spin dynamics of
each population has its unique dependence on temperature,
and how they correlate with time-resolved reflectance
signals.

\end{abstract}


\maketitle

\section{\label{sec:introduction} Introduction}

Spintronics is by now a well established field that aims to
expand the potential of conventional electronics by
exploiting the spin of electrons in addition to their
charge
\cite{wolf2001science,zutic2004rmp,awschalom2009phys}. The
field attracted widespread attention with the proposal of a
spin-based field effect transistor by Datta and Das
\cite{datta1990apl} in 1990. Later, the discovery
\cite{kikkawa1998prl} of exceptionally long spin dephasing
times in bulk \textit{n}-GaAs prompted numerous
investigations of spin dynamics in GaAs based systems. For
double-sided quantum wells (QW) --not doped or with a
two-dimensional electron gas (2DEG)-- various studies with
optical techniques aimed at understanding the more complex
dynamics of 2D electron spin ensembles. The dependence of
spin dephasing times and electron g-factors on parameters
such as confinement energy, carrier density and crystal
orientation has been investigated
\cite{ohno1999prl,brand2002prl,karimov2003prl,averkiev2006prb,holleitner2006prl,yugova2007prb,stich2007prl,koralek2009nature}.

Heterojunction systems, with a strongly asymmetric QW that
contains a high-mobility 2DEG, also have long attracted
considerable interest. These systems provide the highest
values for electron mobility. Further, this material is of
interest since the QW asymmetry results in a strong and
tunable Rashba spin-orbit coupling, that can cancel
Dresselhaus spin-orbit coupling \cite{miller2003prl}, and
this gives access to control over spin relaxation and
dephasing
\cite{koralek2009nature,averkiev1999prb,schliemann2003prl,bernevig2006prl,duckheim2007prb,cheng2007prb,lu2009jap,shen2009jsnm,
liu2009jsnm}. However, most of what is known about electron
spin dynamics in these heterostructures was obtained using
transport measurements
\cite{miller2003prl,potok2002prl,koop2008prl,frolov2009nature}.
Complementing this with magneto-optical pump-probe
measurements brings the advantage that spin dynamics can be
studied with ultra-high time resolution.

We report here on using time-resolved magneto-optical Kerr
rotation (TRKR) \cite{kikkawa1997science} to study the
dynamics of spin ensembles in such a heterojunction QW
system that contains a high-mobility 2DEG. Initial TRKR
studies on such systems suggest that this may also provide
an interesting platform for studying electron-electron
interactions \cite{zhang2008epl,ruan2008prb}. However, it
is not yet well established in which regimes the Kerr
response of 2DEG electrons can be reliably isolated from
other contributions to the Kerr signal. The reason is that
such multilayered GaAs/AlGaAs heterostructures have several
GaAs layers that have an identical value for the gap. This
results in optical readout signals that contain
contributions from (photo-excited) electron populations in
several layers.

For our heterostructure, three different layers can
contribute to the observed Kerr signals from electron
spins: the 2DEG layer, the bulk epitaxial \textit{i}-GaAs
layer that forms the heterojunction and the \textit{i}-GaAs
substrate. The response of these three layers occurs at
nearly the same photon energy, hindering discrimination of
layers by tuning the laser photon energy when using very
short pulses. We exploit, instead, that one can use
differences in the g-factor to isolate the signal from each
population. The g-factor depends on the electron kinetic
energy $E$ (with respect to the bottom of the conduction
band), which can be approximated as
      \begin{equation} \label{eq:gfactorvsE}
           g = g_{0} + \gamma E,
      \end{equation}
where for bulk GaAs \cite{zawadzki1963physlett,yang1993prb}
$g_{0} \approx -0.44$ and $\gamma \approx 6.3 \; {\rm eV
}^{-1}$. For 2DEG electrons these parameters were reported
\cite{yang1993prb} as $g_{0} \approx -0.377$ and $\gamma
\approx 4.5 \; {\rm eV }^{-1}$. Due to band filling by
(photo)electrons, the g-factor thus shifts with an increase
of the (quasi-)Fermi level as well. A related consequence
is that the g-factors shifts towards less negative values
with increasing the degree of confinement in a QW (possibly
with an increased shift from wavefunctions penetrating
AlGaAs barriers). However, it is not obvious that
discrimination by g-factors can be applied without
accounting for electron momentum relaxation and the
intermixing of the populations in different layers
\cite{malajovich2000prl} during the evolution of spin
dynamics. Our measurement technique relies on using a pump
pulse which inserts optically-oriented electron-spin
populations well above the bottom of the conduction band,
and subsequent momentum relaxation and diffusion of
carriers can result in intractable Kerr signals.
Nevertheless, our results show that one can define
situations and regimes (defined by pump intensity) where
the three spin populations can be studied without being
hindered by these effects.

\begin{figure}
\begin{center}
\includegraphics[width=7cm]{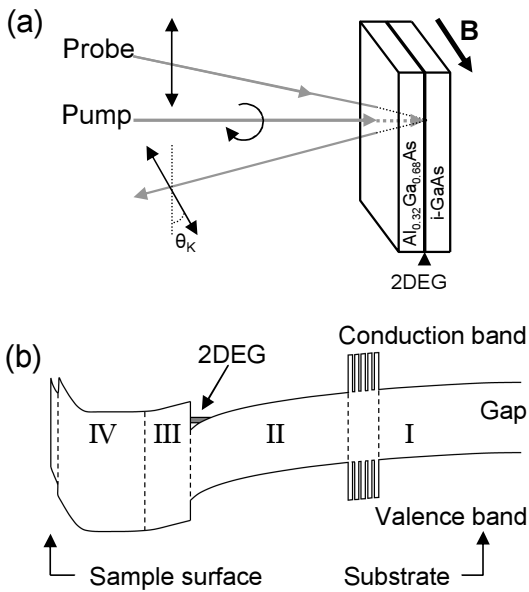}
\end{center}
\caption{(a) Schematic of the setup for time-resolved Kerr
rotation measurements with pump and probe pulses incident
on the 2DEG sample. The pump pulse is circularly polarized
and the probe is linearly polarized. The probe pulses are
incident on the sample at a small angle from the normal
($\sim$2.3$^{\circ}$). The applied magnetic field is
oriented in the plane of the 2DEG. The rotation angle
$\theta_{K}$ of the linear probe polarization is measured
as a function of the time delay between pump and probe
pulses. (b) Schematic of the conduction and valence bands
of the heterojunction system that contains a 2DEG (not to
scale). The composition of the different layers I-IV is
detailed in the main text. At the interface between layers
II and III the conduction bands of the two materials bend,
forming a potential well in layer II where the 2DEG is
formed.} \label{fig:schematic}
\end{figure}

The g-factor of the 2DEG is well separated from that of the
bulk layers, and our results show that the 2DEG population
can be studied as a independent population when the density
of photo-excited carriers does not exceed the 2DEG electron
density from doping. The g-factor for the two bulk layers,
on the other hand, is the same so a different approach is
needed to distinguish the signal originating from each.
Here we use that our heterostructure contains a barrier
that blocks carrier diffusion between the two bulk
\textit{i}-GaAs layers. The effect of this is that the
carrier density in the layer closest to the surface reaches
a higher average value. The effective g-factor for the
uppermost \textit{i}-GaAs layer then acquires a less
negative value (Eq.~\ref{eq:gfactorvsE}) than that of the
underlying substrate. This can be applied in a regime where
the density of photo-excited carriers exceeds the 2DEG
electron density from doping, and under these conditions
the signal from the 2DEG is suppressed by strong band
filling in the 2DEG layer.

In Section~\ref{sec:methods} of this article we discuss the
experimental methods and sample materials that we used.
Results are presented and discussed in
Section~\ref{sec:results}. There, we first compare the Kerr
response of our heterostructure to that of
well-characterized
\cite{kikkawa1998prl,hohage2006apl,schreiber2007arxiv} bulk
\textit{n}-GaAs material that is probed under identical
conditions in our setup, and we highlight the differences.
Next, we show that the response from the heterostructure
can always be described as a superposition of signals from
two independent electron populations, and we confirm that
this is true for both the Kerr signals and for
time-resolved reflectance signals that give insight in
carrier dynamics of the populations. Measurements of spin
dynamics are then analyzed in both the time and frequency
domain, confirming that it is possible to treat the
different electron populations as independent. In order to
confirm that we correctly assign each of the observed
g-factors to a particular population in the heterostructure
we study how the g-factor, spin dephasing time, and carrier
lifetime of each population depend on pump-photon density
and temperature. Section~\ref{sec:conclusion} presents a
summary of the main findings and conclusions.

\section{\label{sec:methods} Methods}

\subsection{Time-resolved probing of electron spins}

In a semiconductor, a laser pulse can promote electrons
from valence-band states to conduction-band states. In a
III-V semiconductor like GaAs, such transitions obey
well-defined selection rules. If the laser pulse is
linearly polarized, the number of photo-excited electrons
with their spin parallel to the beam's propagation
direction equals the number of photo-excited electrons with
their spin antiparallel to the beam's direction. If the
laser pulse is circularly polarized, the number of
photo-excited electrons with spin aligned along the beam's
direction will exceed by 3 to 1 the number of electrons
with spin aligned in the opposite direction (and vice versa
for the opposite light helicity) \cite{meier1984book}. A
circularly polarized laser pulse can therefore generate a
non-equilibrium population of spin polarized electrons in
the conduction band of GaAs. This process is known as
optical spin orientation.

In TRKR experiments, optical spin orientation with an
ultrashort pump pulse is followed by optical spin probing
with a weaker laser pulse with linear polarization that is
reflected on the sample. An unequal filling of the spin-up
and spin-down conduction bands produced by the pump pulse
gives rise to a transient difference in the absorption
coefficient for right and left circularly polarized light
(RCP and LCP respectively). Through Kramers-Kronig relation
this also gives rise to a difference in the refractive
index for RCP and LCP light. Such a difference in the
refractive index gives a rotation of the linear probe
polarization upon reflection on an interface of the sample,
given that the linear polarization is a superposition of
RCP and LCP light. This process is known as Kerr rotation,
and the Kerr rotation angle $\theta_{K}$ (for probing with
a certain photon energy) is proportional to the expectation
value for spin orientation parallel to the propagation
direction of the incident probe beam. The Kerr rotation is
strongest when the probe photon energy is near resonance
with transitions to the states with unequal spin filling,
but its magnitude and sign depend on the detuning with
respect to exact resonance \cite{lee1996apl,kimel2001prb},
and interference effects and additional Faraday rotations
in systems with reflections from multiple interfaces
\cite{salis2006prl}. Note that TRKR signals thus
predominantly reflect the properties of electrons at the
quasi-Fermi level of the photo-induced electron population.

\begin{figure}
\begin{center}
\includegraphics[width=8cm]{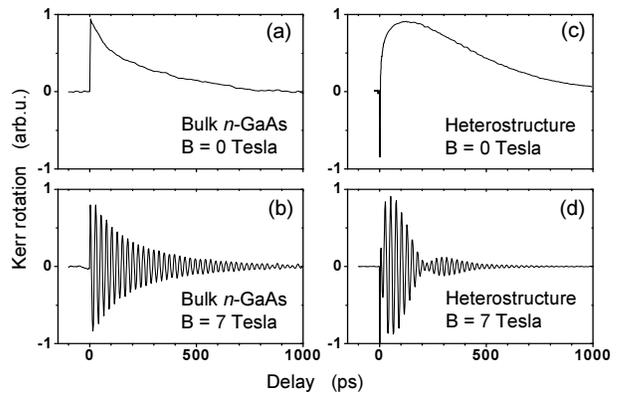}
\end{center}
\caption{Time-resolved Kerr rotation (TRKR) signal at 4.2~K
from bulk \textit{n}-GaAs at 0 Tesla (a) and 7 Tesla (b)
and for the heterostructure containing a two-dimensional
electron gas (2DEG), also at 0 Tesla (c) and 7 Tesla (d).
The data shows considerable differences between the spin
signals from the bulk and 2DEG sample. Most remarkable is
the presence of a node (at $\sim$200~ps) in the Kerr
oscillations measured at 7 Tesla on the 2DEG sample (plot
d). Comparing plots (a) and (c) also clearly shows a slow
increase of the Kerr signal for the 2DEG sample for delays
in the range 0 to 100~ps (also present in the 7 Tesla
data). Data taken with photon density of $\sim90\cdot
10^{11} \; {\rm photons/cm^2}$ per pump pulse.}
\label{fig:bulkand2deg}
\end{figure}

In addition to rotation, the reflected pulse may have
obtained a certain degree of ellipticity, known in this
case as Kerr ellipticity. In our experiments we study the
reflected probe with a polarization bridge with balanced
photodetectors \cite{kikkawa1997science}. The recorded
signal is then proportional to the Kerr rotation, while
Kerr ellipticity and reflectance changes give only third
order corrections. Figure~\ref{fig:schematic}a depicts the
configuration of the experimental setup with pump and probe
beams incident on the sample, and the rotated linear
polarization of a reflected probe pulse. A TRKR trace is
obtained by plotting the Kerr rotation angle $\theta_{K}$
of the reflected probe pulse as a function of the time
delay between the pump and probe pulse (see for example
Fig.~\ref{fig:bulkand2deg}).

In the presence of an in-plane external magnetic field
(perpendicular to the initial orientation of the optically
pumped spins), the injected spins precess around the field
at the Larmor frequency $\omega_{L}$,
   \begin{equation}
      \omega_{L}=g\mu_{B} B/\hbar , \label{eq:larmorfreq}
   \end{equation}
where $g$ is the electron g-factor, $\mu_{B}$ is the Bohr
magneton, $B$ is the magnitude of the applied magnetic
field, and $\hbar$ is the reduced Planck constant. In TRKR
traces this spin precession appears as oscillations of the
Kerr angle at the Larmor frequency
(Fig.~\ref{fig:bulkand2deg}b). Such traces can thus be used
for determining the electron g-factor with
Eq.~\ref{eq:larmorfreq}.

We characterize the transient nature of TRKR traces with a
Kerr signal decay time $\tau_K$. Note that this can be
shorter than the spin dephasing time $T^*_{2}$ for an
electron ensemble. We restrict the use $T^*_{2}$ for the
case that the number of optically oriented electrons in the
probe volume is constant. When this number is decreasing in
time at a rate $1/\tau_e$ due to processes as electron-hole
recombination or diffusion out of the probe volume, the
Kerr signal decays according to $1/\tau_K = 1/T^* _{2} +
1/\tau_e$ (an exception to this rule applies to electron
doped systems, where Kerr signals can live longer than the
recombination time \cite{kikkawa1998prl}). We obtain
$\tau_K$ from measuring TRKR traces at $B=0$ or from
studying the envelope of oscillatory signals at $B > 0$. In
such TRKR traces, hole spins only contribute to the signal
at very short pump-probe delays (up to $\sim 5 \; {\rm
ps}$), because ensembles of hole spins dephase very rapidly
\cite{oestreich1996prb,amand1997prl,linder1998physE,gerlovin2004prb}.
Thus, our TRKR measurements predominantly represent
electron spin dynamics.

The dynamics of photo-excited carriers can be studied
independently from their spin polarization with
time-resolved reflectance $(\Delta R)$ measurement. For
these measurements, a pump pulse with linearly polarized
light is used. In this way, no net spin polarization is
generated. Subsequently, a collinearly polarized probe
pulse is incident on the sample. The pump-induced changes
in sample reflectance are measured by comparing the
intensity of the reflected probe pulse in the presence and
absence of a preceding pump pulse. This method thus probes
the presence of pump-induced electrons and holes at
energies that are resonant with the probe photon energy, as
a function of pump-probe delay. In time, photo-excited
electrons and holes relax to the lowest available states,
and electrons and holes recombine. This brings the sample
reflectance back to its equilibrium value.

\subsection{Sample materials}

Figure~\ref{fig:schematic}b depicts the profile of the
valence band and conduction band of the heterostructure
that we studied. Layer I is a (001) oriented $i$-GaAs
substrate. On this substrate a multilayer buffer consisting
of ten periods of alternating GaAs (5.2~nm) and AlAs
(10.6~nm) layers was grown to smoothen the surface and to
trap unintentional impurities. Layer II, the accumulation
layer, consist of 933.0~nm of undoped GaAs. A spacer layer
(III) is formed by 36.8~nm of undoped
Al$_{0.32}$Ga$_{0.68}$As grown on top of the accumulation
layer. The donor layer (IV) consists of 71.9~nm of Si-doped
Al$_{0.32}$Ga$_{0.68}$As with $\sim 1\cdot 10^{18}$
dopants/cm$^{3}$. The heterostructure is capped with 5.5~nm
of $n$-GaAs. The two AlGaAs layers have a bandgap that is
larger than all the photon energies that we used in our
experiments. Due to quantum confinement the $\sim 5$~nm
GaAs layers of the multilayer buffer and capping layer also
only have optical transitions at energies above the photon
energies that we use.

In this material a 2DEG is formed since dopant electrons
from the donor layer (IV) reduce their energy by relaxing
into the conduction band of he narrow bandgap accumulation
layer (II). These excess carriers are held at the interface
by electrostatic attraction from the ionized donors. The
conduction band of the accumulation layer is pulled below
the Fermi level only for a $\sim$15~nm wide layer near the
interface, which results in the 2DEG at this position
(estimate obtained from solving the coupled 1D
Poisson-Schr\"{o}dinger equations for the heterostructure
\cite{snider1990poisson}). Electron transport experiments
on the 2DEG at 4.2~K, after illumination, gave $2.7 \cdot
10^6 \; {\rm cm^2/Vs}$ for electron mobility, and $4.7
\cdot 10^{11} \; {\rm cm^{-2} }$ for the electron density.
We obtained nearly identical values with optical studies on
our 2DEG material \cite{pugzlys2006jpcm}.

The bulk \textit{n}-GaAs reference material we used had a
doping concentration of $(2.4 \pm 0.2) \cdot 10^{16} \;
{\rm cm^{-3}}$. Several experiments on this material have
shown that at this doping concentration very long spin
dephasing times can be obtained with Kerr studies, and the
dependence of spin dephasing times on various experimental
conditions has been characterized
\cite{kikkawa1998prl,hohage2006apl,schreiber2007arxiv}.

\subsection{Experimental setup}

The measurement setup comprises a magneto-optical flow
cryostat, pulsed laser, polarization optics, and detection
system. Sample temperatures are varied between 4.2 K and
100 K in the flow cryostat. The cryostat is equipped with a
superconducting magnet used to apply fields up to 7 Tesla.
The magnetic field is set parallel to the plane of the
sample while the laser pulses are incident normal to the
sample plane as shown in Fig.~\ref{fig:schematic}a. We used
two different pulsed laser systems. Most of the presented
results were obtained with a cavity-dumped mode-locked
Ti:sapphire laser (further named Laser 1) with 15 fs pulses
corresponding to a fixed spectrum extending from 740~nm to
880~nm, (repetition rates ranging from 4~MHz to 80~MHz).
This allows for wavelength tuning with 10~nm bandpass
interference filters. The central wavelength for both pump
and probe beam can thus be chosen independently. After
filtering, the pulse spectrums have a significant amplitude
in a 19~meV window. The filters are followed by prism
compressors to ensure pulse durations with a full width at
half maximum of approximately 120 fs at the sample. Unless
stated otherwise, we present measurements from using Laser
1 with the pump and probe pulse wavelengths centered at
780~nm and 820~nm respectively. We reproduced most results
in a setup with a tunable Ti:sapphire laser (further named
Laser 2) with $\sim$150~fs pulses at 80~MHz repetition
rate. The spectrum of pulses is wider than
Fourier-transform limited, with significant amplitudes over
a $\sim$15~meV window. With this laser pump and probe
pulses were always centered at the same wavelength.

In each case, two beams derived from the laser are used as
pump and probe with power ratio 24:1 or 4:1. We checked
that all our results were at intensities below the regime
where saturation effects occur: Kerr signals were always
proportional to both the pump and probe intensity, and the
probe was always non-invasive in the sense that the
observed decay times did not depend on probe intensity. We
report the intensity of pump pulses in units of
photons/cm$^2$ per pulse, because it has relevance with
respect to the two-dimensional electron density from
doping. Note that this is the number of photons that is
incident on the sample surface, and that the number is
lower inside the sample due to reflection on each interface
of the heterostructure. Glan-Thompson polarizers on both
pump and probe beam lines are used to produce well defined
polarizations. The pump-probe delay is varied using a
retro-reflector mounted on a stepper-motor driven
translation stage. The two beams are focused on the sample
with a 25 cm focal length spherical mirror which gives
spots with a diameter of approximately 150~${\rm \mu m}$
full-width at base level.

In order to measure the transient Kerr rotation of the
sample, a photo-elastic modulator (PEM) is used to modulate
the polarization of the pump beam between RCP and LCP at a
rate of 50 kHz. Besides improvement of signal-to-noise,
this is crucial for avoiding dynamical nuclear polarization
effects \cite{schreiber2007arxiv}, and it rules out phase
offsets in Kerr signals from interference effects in case
of reflections from multiple interfaces
\cite{salis2006prl}. The probe beam remains linearly
polarized. The reflected probe is analyzed with a
polarization bridge, which decomposes the probe beam into
two orthogonal polarizations with a half-wave plate and a
Wollaston prism, and which detects both these components
with a pair of balanced photodetectors. This bridge is
tuned to give a zero difference signal from the two
detectors for probe pulses with zero Kerr rotation, and
this difference signal is recorded with lock-in detection
at the PEM frequency. For measurements with overlapping
pump and pulse spectrums, we employed a double modulation
technique by adding an optical chopper in the probe beam
(at $f_{C}\sim$1~kHz) and recording a side band at $f_{C}$
from the 50 kHz PEM frequency. In both cases, the signal is
proportional to the Kerr rotation angle $\theta_{K}$.

With this setup it is possible to measure transient
reflectance ($\Delta R$) and TRKR under identical
conditions. For $\Delta R$ measurements the pump and probe
beams have parallel linear polarizations. The reflected
probe beam is now sent directly to a single photodetector,
while the pump beam is modulated using an optical chopper.
Modulations of the reflected probe intensity at the
chopping frequency are then recorded with lock-in
techniques. This gives a signal that is proportional to
pump induced reflectance change of the sample.

\subsection{Relaxation processes and time scales}

Before discussing the experimental data it is useful to
discuss a hierarchy of timescales that is relevant for
interpreting our results. The electron spin dynamics occurs
at time scales of tens to hundreds of picoseconds. For
example, the Larmor precession period of electrons is of
the order of 25~ps at 7~Tesla, and we observe electron spin
dephasing times for the 2DEG in the range of 100~ps. The
duration of the pump and probe pulses (approximately 120
fs) therefore provide sufficient time resolution to resolve
all important spin dynamics. Hole spins dephase within
$\sim$5 picoseconds
\cite{oestreich1996prb,amand1997prl,linder1998physE,gerlovin2004prb}.
Thus, the Kerr signal after a few picoseconds contains
information about electron spins only. Most of the data
shown uses delay steps of $\sim 1$~ps with delays up to 1
ns, which allows us to capture all relevant dynamics of
electron spins.

The timescales of other electron and hole relaxation
processes are very diverse, ranging from a few tens of
femtoseconds to nanoseconds. We typically use a pump pulse
with the photon energy centered at 780~nm, pumping
electrons in GaAs layers at about 63~meV above the bottom
of the conduction band. This leaves a non-equilibrium and
non-thermal carrier distribution in the conduction and
valence bands of the GaAs layers. Within the first hundreds
of femtoseconds, conduction band electrons (and valence
band holes) thermalize into a Fermi-Dirac distribution with
a carrier temperature that is much higher than the lattice
temperature. Subsequently, electrons cool towards the
lattice temperature. The initial cooling of electrons
\cite{leo1988prb,tatham1988solstatelec,kohl1991prb,rosenwaks1993prb,alexandrou1995prb,yucardona1996book},
with excess kinetic energy of more than 10~meV occurs very
rapidly due to rapid momentum relaxation (within 10~ps).
Once the excess kinetic energies drop below $\sim 10$~meV,
further electron cooling slows down to a decay time of
approximately 100~ps depending on electron density. Note
however, that we use probe pulses with about 19~meV
spectral width, with the probe spectrum centered at 820~nm,
just below the bottom of the conduction band of GaAs.
Consequently, most electrons relax into the spectral window
of the probe within 10~ps, and the final cooling stage in
the time interval from 10 to 100~ps has little influence on
our Kerr and reflectance signals. Thus, the fast carrier
thermalization and initial cooling allows for pumping the
sample with high-energy photons (convenient for blocking of
pump light that scatters into the probe channel by spectral
filtering, and it ensures that the penetration depth of
pump light for the top GaAs layers is less than $1 \; {\rm
\mu m}$). Right after a pump pulse, there will be a steep
gradient in the photo-electron density along the direction
of the pump over the length of the penetration depth. This
equilibrates by electron diffusion along the beam
direction. In the accumulation layer this takes place
within the first 50~ps for high pump intensities, but may
occur much slower \cite{salis2006prl} for low pump
intensities. For our spot sizes, electron diffusion into
directions perpendicular to the laser beam propagation does
not significantly change the electron density profile.

Other carrier processes that take place in timescales
between 1~ps and 1~ns are exciton formation and carrier
recombination. Within the first hundred picoseconds
electrons and holes form excitons. This implies a further
reduction of the initial energy of the photo-excited
electrons by a few meV, but this occurs within the spectral
window of our probe, and these processes do therefore not
strongly influence our $\Delta R$ and TRKR traces.
Electron-hole recombination, on the other hand, is directly
visible in $\Delta R$ measurements. Obviously,
recombination also results in loss of Kerr signal,
especially in undoped samples or under conditions where the
photo-excited electron density exceeds the electron density
due to doping \cite{kikkawa1997science}. With time- and
spectrally-resolved  photoluminescence, we measured that
the exciton recombination time for our 2DEG sample at 4.2 K
is greater than 2~ns in the accumulation layer, while it is
approximately 360~ps for the substrate
\cite{pugzlys2006jpcm}. Both of these timescales become
shorter with increasing pump intensity.  As a function of
temperature the 2DEG recombination time increases, while it
decreases for bulk GaAs layers and $n$-GaAs.

\section{\label{sec:results} Results and discussion}

\subsection{\label{subsec:GaAsKerr} Kerr signals from a GaAs/AlGaAs heterojunction system}

In order to highlight the Kerr response that is
characteristic for the heterostructure, we compare it to
Kerr measurements on the bulk \textit{n}-GaAs sample that
were obtained under identical conditions.
Figure~\ref{fig:bulkand2deg} shows in the left column TRKR
signals from the bulk \textit{n}-GaAs sample. Its Kerr
signal at 0 Tesla closely resembles a mono-exponential
decay. The signal at 7 Tesla shows oscillations at the
Larmor precession frequency for electrons in GaAs (g-factor
is $|g| \approx 0.44$ \cite{gfactordriftvalue}), with an
envelope that shows again mono-exponential decay
\cite{ShortnGaAs}. The transient TRKR signals from the
heterostructure, on the other hand, are more complex. At 7
Tesla (Fig.~\ref{fig:bulkand2deg}d), the Kerr response
shows a node (at $\sim$200~ps), indicating a beating
between different precession frequencies. Additionally, the
envelope shows a slow increase in the delay range from 0 to
100~ps, which is also observed at 0 Tesla
(Fig.~\ref{fig:bulkand2deg}c). The \textit{in-situ}
comparison with bulk $n$-GaAs material is important for
concluding that the observed beatings in TRKR traces are
characteristic for the heterostructure, given that beatings
can also appear in Kerr oscillations from $n$-GaAs under
certain measurement conditions \cite{schreiber2007arxiv}.

The heterostructure data in Figs.~\ref{fig:bulkand2deg}c,d
also show a very sharp spike at very short delays, with a
decay time of about 3~ps. It is also present in several
other TRKR traces from the heterostructure that we will
present and it was also observed with the $n$-GaAs sample
at weaker pump intensities. It is longer than the duration
of pump-probe overlap, and therefore not an optically
induced Stark effect \cite{kimel2001prb}. Instead, it
occurs at a timescale that is consistent with hole spin
dephasing, and hot-electron momentum relaxation. This part
of the Kerr signal is therefore possibly influenced by
these effects. We further focus on the Kerr signals at
delays of 5~ps and longer, and do not include this spike in
further analysis.

\begin{figure}
\begin{center}
\includegraphics[width=8cm]{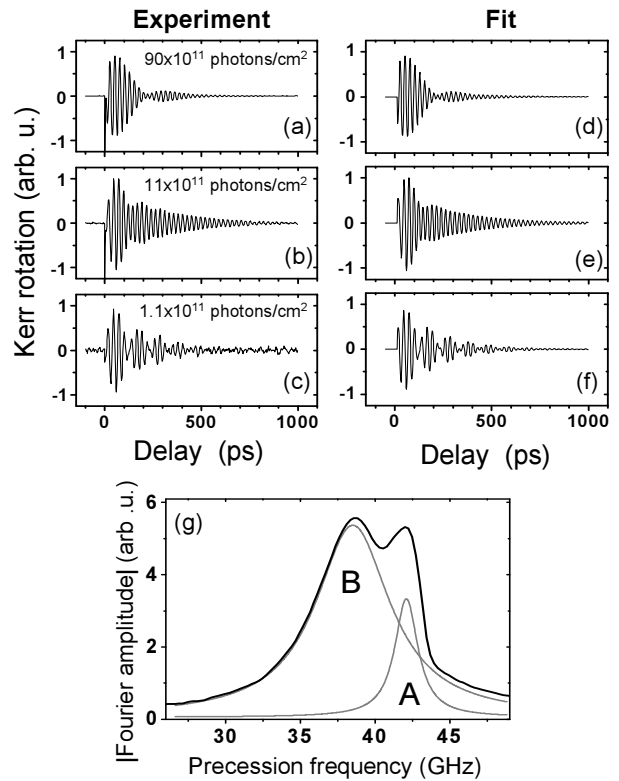}
\end{center}
\caption{Time-resolved Kerr rotation signal of the 2DEG
sample for various pump photon densities. The left column,
plots (a)-(c), show the sample response at photon densities
per pump pulse as labeled. The pump-photon density strongly
influences the beating pattern in the Kerr oscillations.
The right column, plots (d)-(f), show fits to the data in
(a)-(c) using Eq.~\ref{eq:fitfunction} (see main text for
details). Plot (g) shows a Fourier transform (black line)
of the TRKR data of plot (a), showing to distinct
precession frequencies (with labels for populations A and B
as in Fig.~\ref{fig:gfactorvsfluence}). Good fits to the
spectrum can be obtained using a superposition of two
Lorentzians (gray lines). Data taken at 7 Tesla and 4.2~K.}
\label{fig:TRKRvsfluence}
\end{figure}

Figures~\ref{fig:TRKRvsfluence}a-c show that the beatings
and slow onset in the heterostructure Kerr signals appear
for a wide range of pump intensities. However, the exact
appearance clearly depends on pump intensity. In what
follows we will show that all three traces in
Figs.~\ref{fig:TRKRvsfluence}a-c can be described as the
superposition of two oscillatory signals with
mono-exponential decay remarkably well. Each of the two
contributions then results from a different electron
population, each with its own decay time and effective
g-factor. At the highest pump intensities
(Fig.~\ref{fig:TRKRvsfluence}a), the signal is dominated by
two bulk $i$-GaAs electron populations, one in the in the
substrate (layer I in Fig.~\ref{fig:schematic}b) and one in
the accumulation layer (II). At the lowest pump intensity,
(Fig.~\ref{fig:TRKRvsfluence}c) the signal is dominated by
a population in the 2DEG quantum well, and a second one
with bulk $i$-GaAs characteristics.

Figure~\ref{fig:TRKRvsfluence}g shows the Fourier transform of
the trace in Fig.~\ref{fig:TRKRvsfluence}a. The amplitude
spectrum shows two distinct peaks. It can be fit very well
with two Lorentzians, and the same holds for the Fourier
transform of the traces in Figs.~\ref{fig:TRKRvsfluence}b,c.
This suggests that one can fit the traces in
Fig.~\ref{fig:TRKRvsfluence}a-c in the time-domain with a
superposition of two mono-exponentially decaying cosine
functions. That also builds on observations in $n$-GaAs
where TRKR signals from single electron population with a
single g-factor (see also Fig.~\ref{fig:bulkand2deg}b) can be
described by a mono-exponentially decaying cosine function
\cite{kikkawa1998prl}. For this fitting we use
   \begin{eqnarray}\label{eq:fitfunction}
      \theta_{K}=A_{1}\exp({-t/\tau_{K1}})\cos(2\pi\omega_{L1}t+\phi_{1})\notag\\
      +A_{2}\exp({-t/\tau_{K2}})\cos(2\pi\omega_{L2}t+\phi_{2}).
   \end{eqnarray}
Here $t$ is the pump-probe delay, $\tau_{K1}$ $(\tau_{K2})$
and $A_{1}$ $(A_{2})$ are, respectively, the Kerr signal
decay time and the Kerr rotation amplitude for each
population. Similarly, $\omega_{L1}$ and $\omega_{L2}$ are
the Larmor frequencies and $\phi_{1}$ and $\phi_{2}$ are
apparent initial phases \cite{phases} for spin precession.
For fitting we only use data for $t>5$~ps.

The fitting results of applying this to the experimental
traces of Fig.~\ref{fig:TRKRvsfluence}a-c are presented in
Fig.~\ref{fig:TRKRvsfluence}d-f. The excellent fits
demonstrate that the total signal can be indeed described
as being composed of two distinct contributions from two
different populations. We find that we can obtain very good
fits assuming two populations only for all pump
intensities. A similar approach that assumes three
populations does not improve the fits significantly. This
is a remarkable result, given that the Kerr response of our
heterostructure does not necessarily split into two
distinct contributions. The continuous character of the
accumulation layer (II) from quantum well to bulk epilayer,
followed by a bulk substrate, could give more complex
signals with signatures of relaxation processes and mixing
effects between populations \cite{malajovich2000prl}. We
find that the sign of the amplitudes $A_1$ and $A_2$ of the
two contributions is always opposite (for $\phi_1$ and
$\phi_2$ close to zero) which corresponds to Kerr rotations
of opposite sign at $t=0$. Consequently, the slow signal
increase at early delays and the beatings in our Kerr
oscillations have a common origin. The sign of the Kerr
response depends on detuning from resonance, but we
predominantly probe slightly red-detuned for all
populations. The opposite sign is therefore most likely
from interference effects that play a role for the sign of
Kerr response with reflections from multiple interfaces
\cite{salis2006prl}.

Further evidence that these excellent fits in the time
domain indeed reflect that the physics that underlies our
Kerr signals is that of two independent populations, each
with a different but (nearly \cite{gfactordriftvalue})
constant g-factor, comes from Fourier analysis of the
traces. Scenario's that include significant spreads or
strongly drifting g-factors due to momentum relaxation can
be ruled out based on the phase spectrum of Fourier
transformed TRKR signals. In addition, direct measurements
of the momentum-relaxation dynamics support this
interpretation.

\begin{figure}
\begin{center}
\includegraphics[width=6.5cm]{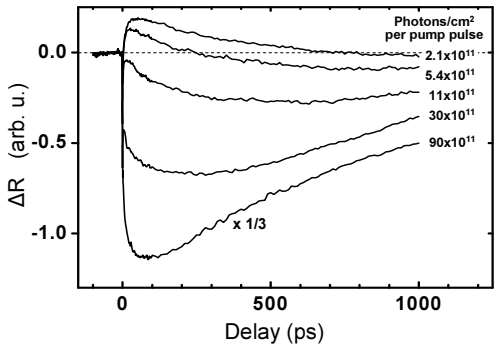}
\end{center}
\caption{Transient reflectance $(\Delta R)$ traces from the
2DEG sample at different pump-photon densities.
Data taken at 4.2 K and 0 Tesla.}
\label{fig:Transrefvsfluence}
\end{figure}

We applied time-resolved reflectance measurements for these
studies of the momentum-relaxation dynamics. It provides a
direct measure for the momentum relaxation of electron
populations into the spectral window of the probe and
subsequent electron-hole recombination.
Figure~\ref{fig:Transrefvsfluence} shows $\Delta R$ as a
function of pump-probe delay measured on the
heterostructure sample using different pump intensities. At
high pump-photon densities, $\Delta R$ first reaches a
negative peak before decaying back to zero. At low
pump-photon densities, the signal first shows a positive
$\Delta R$ but at later delays $\Delta R$ obtains a
negative value with only very slow decay back to zero.

Also here we find that we can describe the full set of
$\Delta R$ traces for $t>10$~ps as the sum of two
mono-exponentially decaying responses of opposite sign: one
response with positive $\Delta R$ that is short lived and
another response with negative $\Delta R$ that is long
lived. In addition, there is for each trace during $0<t
\lesssim 10$~ps an onset proportional to
$1-\exp(-t/\tau_{mr})$ with a decay $\tau_{mr}\lesssim
10$~ps. This is consistent with the initial momentum
relaxation of pumped electrons into the spectral window of
the probe. For $t > 10 \; {\rm ps}$, we obtain excellent
fits with an approach analogous to fitting TRKR signals
with Eq.~\ref{eq:fitfunction} (but with the cosine factors
set to 1). The difference in shape for traces recorded at
different pump intensities is then predominantly due a
shift in the relative weight of the two contributions.

At the lowest pump intensities, the signal is first
dominated by a short lived ($\sim 360$~ps) positive
contribution to $\Delta R$. At later delays a longer lived
(2-3~ns) negative contribution dominates the signal. This
is consistent with the photoluminescence
results\cite{pugzlys2006jpcm} if we assign the positive
short-lived contribution to the substrate, and the negative
long-lived contribution to the 2DEG layer. Going to higher
pump intensities, the relative weight of the negative
contribution (ratio $|A_2/A_1|$, with $A_2$ the amplitude
of the negative contribution) increases from $\sim 0.5$ to
$\sim 5$. This is consistent with an increased importance
for the accumulation layer (II) that can be expected. In
addition, the decay time for negative long-lived population
decreases to $\sim 1$~ns. Note, however, that at high pump
intensity there is significant band-filling in the
accumulation layer (II), so it should now be considered the
recombination time for an electron population that is
extended throughout this layer instead of a 2DEG
population. At the high pump intensities, the onset of the
strong negative contribution (with onset proportional to
$1-\exp(-t/\tau_{mr})$) curves over into the weak positive
contribution with a decay time of $\lesssim$360~ps, which
makes it impossible to reliably fit this latter time scale
for high-pump intensities. When increasing the temperature
to 100 K, the life time of the negative contribution
gradually shortens to $\sim$100~ps, while the life time of
the positive contribution grows to $\sim$8~ns.

Thus, we find that both TRKR signals and $\Delta R$ signals
can be described very accurately as the superimposed
response of two different electron populations. Moreover,
the relative amplitudes $A_1$ and $A_2$ of populations that
are observed with TRKR correlate with those observed in
$\Delta R$ signals (further discussed at
Fig.~\ref{fig:gfactorvsfluence} below). The fact that the
signal or signal-envelope for each of the two contributions
is well described by mono-exponential decay confirms that
photo-excited electrons cool within about 10~ps to a
temperature where all electrons are at energies within the
spectral window of the probe. Further cooling only causes
small changes in the average value and spread of g-factors
and are masked by the relatively short decay times of the
Kerr signals. Further, we observe that for the substrate
contribution to the signals, the decay times for the
reflectance signals and Kerr signals are comparable. This
means that the loss of Kerr signals is here for a
significant part due to electron-hole recombination.

\begin{figure}
\begin{center}
\includegraphics[width=8cm]{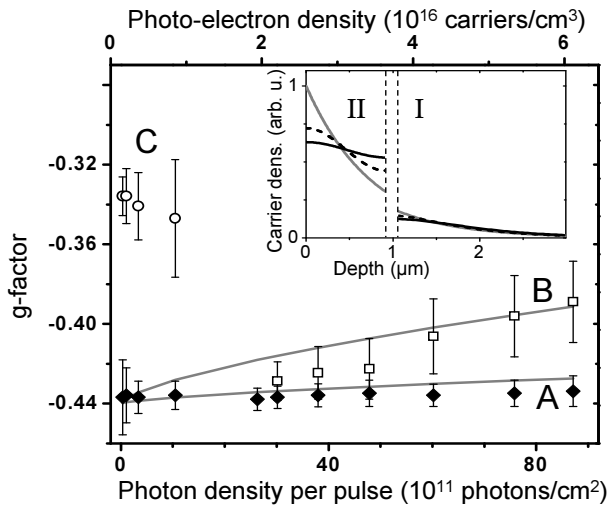}
\end{center}
\caption{The g-factors of spin populations observed in the
2DEG sample as a function of pump-pulse photon density
(bottom axis) and the estimated photo-excited electron
density in the accumulation layer (top axis). At high
photon densities two spin precession modes can be resolved
with g-factors $|g| \approx 0.44$ (solid diamonds,
population A) and $|g| \approx 0.39$ (open squares,
population B). We ascribe these g-factors to electrons in
the \textit{i}-GaAs substrate (layer I) and electrons in
the accumulation layer (layer II) respectively. The gray
lines through the data points of A en B are calculated
g-factor values (see main text for details). A third
precession mode (open circles, population C) is observed at
the lowest pump-photon densities and corresponds to
electrons in the 2DEG. Data obtained under conditions as in
Fig.~\ref{fig:TRKRvsfluence}. The inset shows the
calculated density of photo-excited electrons in the
accumulation layer (layer II) and in the substrate (layer
I) as a function of depth from the top of the accumulation
layer. The dashed vertical lines at 0.933 $\mu$m show the
position of the multilayer buffer. The gray line shows the
calculated electron density immediately after photo
excitation with a 780~nm pump pulse. The dashed black line
(solid black line) shows the electron density 20~ps (50~ps)
after photo excitation.} \label{fig:gfactorvsfluence}
\end{figure}

Although the TRKR traces from the heterostructure can
always be fit assuming two spin populations, a total of
three different spin populations show up in the TRKR
measurements when these are studied for a range of pump
pulse intensities. From such data, we analyze the
precession frequencies of the different spin populations,
both with Fourier analysis as in
Fig.~\ref{fig:TRKRvsfluence}g, and fitting in the time
domain (Eq.~\ref{eq:fitfunction}).
Figure~\ref{fig:gfactorvsfluence} shows which g-factors are
observed as a function of pump-pulse intensity. Two spin
populations show up at high pump-photon density (labeled A
and B) a third population is only observed at low
pump-photon density (labeled C). The ratio of contributions
$|A_{C}/A_{A}|$, going over into $|A_{B}/A_{A}|$, increases
from $\lesssim 1$ to $\sim 5$ with increasing pump
intensity. It thus closely follows the ratio
$|A_{2}/A_{1}|$ discussed for
Fig.~\ref{fig:Transrefvsfluence}. This  provides the first
evidence that signal contributions B and C come from the
accumulation layer, and A from the substrate. In the next
Sections (\ref{subsec:iGaAsKerr} and \ref{subsec:2degKerr})
we further discuss the origin of these three spin
populations. We also analyze the physics that results in
the fact that these populations behave independently, and
the spin dynamics of each population. The error bars
indicate here the uncertainty in position for the two
Lorentzian peak shapes that we fit on the spectrums, and
thus also give an indication for the Kerr decay times for
each contribution (the smallest error bars corresponding to
$\sim$250~ps, the largest to $\sim$70~ps).

\subsection{\label{subsec:iGaAsKerr} Kerr signals from $i$-GaAs populations}

The value of the g-factor of each population gives a strong
indication of the medium that the electrons populate. The
g-factors of the different semiconductor materials that
make up the heterostructure are well known. The g-factor of
electrons in bulk GaAs is approximately $-0.44$
\cite{weisbuch1977prb}, and electrons in heterojunction
2DEGs and 15 to 20~nm wide GaAs QWs have a g-factor of
about -0.36
\cite{snelling1991prb,snelling1992prb,jiang2001prb,meisels2005semiscitech,yugova2007prb}.
For electrons in the 5.5~nm wide GaAs layers $g \approx
+0.15$, and for \cite{hannak1975solstatcomm} bulk
Al$_{0.32}$Ga$_{0.68}$As layers $g \approx +0.5$
\cite{weisbuch1977prb} (but electron populations in these
latter two layers are not interacting with the photon
energies that we use).

The value of the g-factor of population A is $|g| \approx
0.44$ at all pump-photon densities. We therefore associate
population A with electrons in bulk GaAs. Measurements of
the temperature dependence of the g-factor of population A,
discussed below, also indicate that these electrons
populate a bulk layer of GaAs. Time-resolved Kerr rotation
measurements cannot directly determine the sign of
g-factors. However, the g-factor of GaAs is known to be
negative \cite{weisbuch1977prb} so we assign a negative
value to the g-factor of population A.

The g-factor of population B shows a strong dependence on
pump-photon density. At the highest pump-photon densities
the g-factor of population B is $|g| \approx 0.39$ (while
the signal contribution B is here about 5 times stronger
than A with its Kerr signal decay time  as short as
$\sim$70~ps, see also the broad peak in
Fig.~\ref{fig:TRKRvsfluence}g). As the photon density is
reduced, the g-factor of population B approaches the value
of $|g| \approx 0.44$. This shows that this population also
corresponds to electrons in bulk GaAs, and we again assign
a negative value to its g-factor. There are two bulk
\textit{i}-GaAs layers in the heterostructure: the
accumulation layer (II, see Fig.~\ref{fig:schematic}) and
the substrate (I). The analysis in the next paragraphs
demonstrates that population A corresponds to electrons in
layer I, and that population B corresponds to electrons in
layer II. Population C shows up in the TRKR measurements
only at low pump-photon densities, and corresponds to 2DEG
electrons, as will be further discussed in
Section~\ref{subsec:2degKerr}.

At high pump-photon densities, the response from the two
\textit{i}-GaAs layers can be distinguished because
different average electron densities are excited in each
layer by the pump pulse, which results in different
g-factors (Eq.~\ref{eq:gfactorvsE}). In the following
discussion we will show how different average electron
densities are obtained thanks to a pump beam with short
penetration depth and the multilayer buffer between layer I
and layer II. At 4.2~K the absorption coefficient of 780~nm
light in GaAs gives a penetration depth of approximately
0.77~$\mu$m \cite{sturge1962pr}. The density of
photo-excited electrons that is present immediately after a
pump pulse decays exponentially as a function of depth into
the GaAs layers over this length scale. Thus, most of the
incident pump photons are absorbed in the accumulation
layer (II), giving a much higher concentration of
photo-excited electrons in this layer than in the substrate
(I). The gray line in the inset in
Fig.~\ref{fig:gfactorvsfluence} shows the density of
photo-excited electrons as a function of depth immediately
after the absorption of a pump pulse. This strong gradient
in the photo-electron density will rapidly equilibrate due
to diffusion. The inset also shows the calculated electron
density as a function of depth 20~ps and 50~ps later. We
used here a bulk diffusion constant
\cite{wolfe1970prb,pugzlys2006jpcm} of 30~cm$^2$/s, but the
results of the model that we present below are quite
independent from the exact value of diffusion constant that
is used. In this calculation, the influence of the
space-charge potential in the accumulation layer is
ignored. This is justified at high pump-photon densities
because the space-charge potential is screened by the
photo-excited carriers. In addition, carrier-recombination
effects are ignored since these occur at timescales longer
than the considered diffusion times. Note that the
penetration depth for part of the probe spectrum (centered
at 820~nm) is considerably longer than that of the pump,
such that the Kerr response of the deepest photo-excited
electrons is not strongly attenuated with respect to
electrons near the surface. In addition, the reflection on
the multilayer buffer layer ensures that there is an
enhanced Kerr response for electrons in its direct
vicinity.

The dashed vertical line in the inset in
Fig.~\ref{fig:gfactorvsfluence} at 0.933~$\mu$m represents
the position of the multilayer buffer, which acts as a
barrier that blocks electron diffusion. The calculated
electron density profiles at 20~ps and 50~ps show that the
electron density, and thus the quasi-Fermi level in the two
layers, becomes discontinuous at the multilayer buffer. The
accumulation layer reaches a much higher average electron
density than the substrate (see also top axis of
Fig.~\ref{fig:gfactorvsfluence} for estimated values), and
thereby a much higher value for the  of the quasi-Fermi
level that is established after tens of picoseconds when
most electrons are in the lowest available conduction band
states (besides spin relaxation) due to cooling and
diffusion. The average electron g-factor of a population
that is observed in TRKR signals depends on the quasi-Fermi
level as in Eq.~\ref{eq:gfactorvsE}. We used this to
calculate the expected g-factors of electrons near the
quasi-Fermi level for the population in the accumulation
layer (II) and the substrate (I), without any adjustable
parameters \cite{oestreich1996prb}. When calculating a
photo-electron density from pump-photon density that is
incident on the heterostructure surface we accounted for an
estimate of the reflection on each interface in the
heterostructure. These calculated g-factors are also
plotted in Fig.~\ref{fig:gfactorvsfluence}. The good
agreement between the measured and calculated g-factors for
populations A and B confirms that the two populations are
indeed electrons in the accumulation layer and the
substrate.

The above discussion shows that the ability to distinguish
the two populations relies on the short penetration depth
for the pump pulse. Our results did not change
significantly when using 800~nm pump and 820~nm probe
photons (also showing that the initial rapid momentum
relaxation does not result in significant spin dephasing).
However, when using 820~nm photons for both pump and probe
with Laser 1, we did no longer observe the beatings in our
TRKR oscillations. Under these conditions, the relative
contribution from the population in the substrate is much
higher due to deeper penetration of pump light into GaAs
(the penetration depth increases beyond 3~$\mu$m when the
wavelength increases beyond 820~nm \cite{sturge1962pr}).
This prohibits an analysis of the contribution to the Kerr
signal from the top GaAs layers of the heterostructure. We
could, nevertheless, fully reproduce our results in a
mono-color experiment, explored with the tunable Laser 2.
Also here, we had to ensure that the full spectrum of laser
pulses had a short penetration depth. We could get results
with a strong contribution from the accumulation layer to
the Kerr response (both at high and low pump intensities)
by setting the central laser wavelength at least
$\sim$20~meV above the bottom of the conduction band.
Momentum relaxation brings electrons again rapidly near the
bottom of the conduction band, and probing then occurs
predominantly with the low-energy wing of our pulse
spectrum. When tuning the central wavelength above
$\sim$810~nm, the Kerr response was again fully dominated
by electrons in the substrate.

\begin{figure}
\begin{center}
\includegraphics[width=8cm]{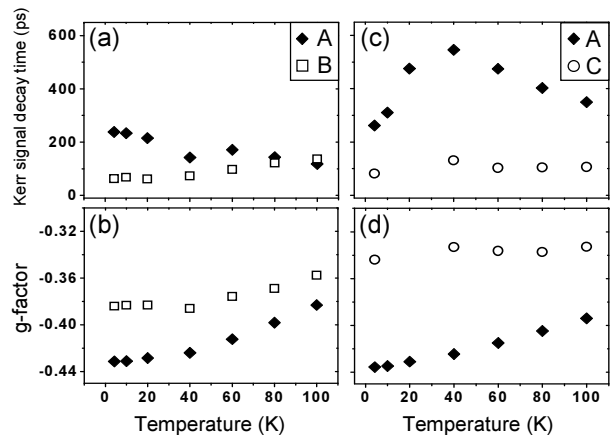}
\end{center}
\caption{Kerr signal decay times (top row) and g-factors
(bottom row) as a function of temperature, measured at pump
photon densities of $\sim100\cdot 10^{11}\: {\rm
photons/cm^2}$ (left column) and $\sim3.6\cdot 10^{11}\:
{\rm photons/cm^2}$ (right column) in the heterostructure
sample. Electron populations A, B, and C (see also
Fig.~\ref{fig:gfactorvsfluence}) are indicated by open
squares, solid diamonds, and open circles respectively. The
data points were obtained from fits with
Eq.~\ref{eq:fitfunction} to TRKR measurements taken at 7
Tesla.} \label{fig:fitparamvstemp}
\end{figure}

As an additional check for our interpretation in this
regime with high pump-photon densities we studied how the
Kerr signals depend on temperature $T$ (presented in
Fig.~\ref{fig:fitparamvstemp}) and magnetic field $B$ (only
discussed). The Kerr signal decay times for populations A
and B as a function of temperature are presented in
Fig.~\ref{fig:fitparamvstemp}a, and the corresponding
g-factors in Fig.~\ref{fig:fitparamvstemp}b. These results
were obtained from fitting Eq.~\ref{eq:fitfunction} to TRKR
measurements taken at $\sim100\cdot 10^{11} \: {\rm
photons/cm^2}$ per pump pulse. The temperature dependence
of the g-factors of populations A and B is in agreement
with observations by Oestreich \textit{et al.} on GaAs bulk
samples \cite{oestreich1996prb}, confirming that
populations A and B correspond to electrons with bulk GaAs
characteristics. The g-factor of population B increases
with temperature in a manner similar to population A.
However, with increasing temperature the difference
decreases. This is consistent with a broadening of the
quasi-Fermi level with temperature, which increases the
average electron kinetic energy $E$ and (by
Eq.~\ref{eq:gfactorvsE}) brings the g-factor values closer
to zero. At the highest temperatures, the broadening of the
quasi-Fermi level starts to become larger than the
difference in quasi-Fermi level for populations A and B,
which results in a smaller difference between the two
g-factor values.

The Kerr signal decay times $\tau_{KA}$ and $\tau_{KB}$ of
populations A and B show opposite trends as a function of
temperature. At 4.2 K the decay time of population A is
approximately 240 ps. As the temperature is raised this
value decreases monotonically down to $\sim$120~ps. The
signal decay time of population B, on the other hand,
increases from $\sim$70~ps at 4.2~K to $\sim$140~ps at
100~K. We analyzed that several mechanism contribute to
these Kerr signal decay times, and full understanding goes
beyond the scope of this article. Instead, we will only
discuss a few trends and typical values. The following
three mechanism contribute to the Kerr signal decay rates
that we observe:

\noindent \textit{i}) Electron-hole recombination. This
directly contributes to loss of Kerr signals at the
electron-hole recombination rate.

\noindent \textit{ii}) Precessional dephasing due to a
spread $\Delta g$ in g-factor values. At high pump-photon
densities, band filling gives that the spin-oriented
electrons have a spread in electron kinetic energy $E$,
which (by Eq.~\ref{eq:gfactorvsE}) directly results in a
spread $\Delta g$. This results in a dephasing rate
\cite{kikkawa1998prl} $1/T_2^{*} = \Delta g \mu_B B /
\sqrt{2} \hbar$.

\noindent \textit{iii}) The D'Yakonov-Perel' (DP) spin
dephasing mechanism
\cite{dyakonov1971jetp,dyakonov1986sovps}, as a result of
the cubic Dresselhaus spin-orbit coupling in GaAs: The
spins of electrons in motion experience a $k$-vector
dependent effective magnetic field. Consequently, random
momentum scattering (at a rate $1 / \tau_{p}$) randomizes
spin states by precession. At low temperatures and high
electron densities, this effect contributes to spin
dephasing at a rate (for a review see
Ref.~\onlinecite{zutic2004rmp}) $1/T_2^{*} \propto \tau_{p}
\gamma_{cub}^2 n_{3D}^2$, where $\gamma_{cub}$ is the cubic
Dresselhaus constant, and $n_{3D}$ the bulk
(photo-)electron density. The rate of this effect typically
decreases with increasing magnetic field if $\tau_p$ is
long enough to yield cyclotron motion. We estimate the DP
rates with a numerical Monte-Carlo approach, that we
described in detail in Ref.~\onlinecite{koop2008arxiv}.

In our experiments, these contributions are often of
similar magnitude, but the weights shift with temperature,
field and electron density. At 4.2~K and 7~T, and for
pumping at $\sim100 \cdot 10^{11} \: {\rm photons/cm^2}$,
the estimate for the precessional dephasing rate by $\Delta
g$ gives about (1~ns)$^{-1}$ for population A, and
(150~ps)$^{-1}$ for population B, thus not fully explaining
the observed decay rates. In addition, both $\tau_{KA}$ and
$\tau_{KB}$ gradually increase by a factor $\sim$3 when
lowering $B$ towards zero. This dependence is weaker than
$1/B$ for both $\tau_{KA}$ and $\tau_{KB}$, confirming that
a spread in $\Delta g$ is not fully responsible for the
observed decay. The electron-hole recombination time for
population A is about $\lesssim 300$~ps at 4.2~K
(Section~\ref{subsec:GaAsKerr}), and was observed to
decrease to about 100~ps at 100~K. It thereby provides the
dominant rate contributing to the decay of A, and also
explains the temperature dependence of $\tau_{KA}$ in
Fig.~\ref{fig:fitparamvstemp}a. This is supported by the
fact that the $n$-GaAs reference sample (measured under
identical conditions) shows here very similar decay times
and temperature dependence as population A.

As said, the dependence of $\tau_{KB}$ on temperature is
opposite to that of $\tau_{KA}$. For population B (in the
accumulation layer), the electron-hole recombination time
is much longer than for A (Section~\ref{subsec:GaAsKerr}),
and increasing from $\sim$1~ns to $\sim$8~ns when
increasing $T$ to 100~K. Thus, electron-hole recombination
barely contributes to the observed $\tau_{KB}$ values.
However, due to the much higher electron density for B, the
DP mechanism now probably contributes significantly. The
Monte-Carlo simulation (for 0~K, 7~T, $n_{3D} = 7 \cdot
10^{16}$~cm$^{-3}$, and the typical value
\cite{miller2003prl} $\gamma_{cub} = 30 \cdot 10^{16} \:
\rm{eV\,nm^{3}}$) give a DP rate of (200~ps)$^{-1}$ if we
assume a high mobility of $4 \cdot 10^5 \; \rm{cm^2 / Vs}$
(giving $\tau_p = 150 \; \rm{ps}$). Such high mobility
values were indeed directly observed in transient grating
experiments on our sample material \cite{pugzlys2006jpcm}
at these high photon-electron densities, and due to the
fact that population B is in a pure undoped layer of
epitaxial quality. We can also explain the temperature
dependence of B in Fig.~\ref{fig:fitparamvstemp}a when
making this assumption. The observed momentum scattering
rate $1/ \tau_p$ increases by a factor $\sim$30 in
proportion to $T$ in the range 10~K to 100~K
\cite{pugzlys2006jpcm} due to acoustic phonon scattering
\cite{mendez1984apl}. The shortening of $\tau_p$ towards
100~K thus results in a reduced DP dephasing rate with
increasing temperature. A moderate increase of $\tau_{KA}$
and constant behavior of $\tau_{KB}$ when lowering the pump
intensity is consistent with the rate contributions that we
discussed here.

\subsection{\label{subsec:2degKerr} Kerr signals from the 2DEG population}

Figure~\ref{fig:gfactorvsfluence} shows that population C
with $|g|\approx 0.34$ (besides a population with $|g|
\approx 0.44$) can be observed when the pump-photon density
drops below $\sim$$10\cdot 10^{11}\: {\rm photons/cm^2}$
per pulse. This crossover occurs around the 2DEG electron
density due to doping, and this provides another indication
that population C resides in the 2DEG quantum well. We
assume again that the g-factor of this population is
negative, and its value ($\sim-0.34$) is then indeed
consistent with the g-factor of electrons in heterojunction
2DEG systems and 15 to 20~nm wide GaAs QWs
\cite{snelling1991prb,jiang2001prb,meisels2005semiscitech}.
Other electrons populations that could be considered to
give rise to the signal of population C can be convincingly
rejected based on the g-factor values that we discussed
before (Section~\ref{subsec:iGaAsKerr}). One can also rule
out that population C corresponds to electrons or excitons
trapped at impurities or defects in bulk GaAs, since a
g-factor of this value has not been observed before by us
or by others in bulk GaAs samples.

We further remark that the g-factor values $|g| \approx
0.44$ and $|g| \approx 0.34$ rule out the interpretation
that observing two populations at these low pump
intensities results from simultaneous probing of electrons
in two different subbands of the heterojunction QW.
Electrons in excited subbands have a g-factor that is
closer to zero than the g-factor $g \approx -0.34$ of
electrons in the lowest subband \cite{winkler2003book}.
Hence, the signal contribution with $|g| \approx 0.44$
cannot result from 2DEG electrons. Such an interpretation
would also be inconsistent with the observed decay times
for the two populations in both TRKR traces and $\Delta R$
traces. Instead, the population with $|g| \approx 0.44$
must be due to electrons with bulk GaAs characteristics
around the multilayer buffer, and we label it again with A.
However, for these low pump intensities we were not able to
determine whether this signal contribution is dominated by
electrons just above (in layer II) or below (in layer I)
the multilayer buffer (further discussed below).

We carried out two additional experiments which confirm
that population C corresponds to 2DEG electrons. In the
first experiment, we performed TRKR measurements on a
heterostructure where the 2DEG had been removed by wet
etching to a depth of about 120~nm. On such material, one
can still observe populations A and B at high pump-photon
densities. At low pump-photon densities, however,
population C is no longer observed, and Kerr signals only
contain a contribution with $|g| \approx 0.44$.

\begin{figure}
\begin{center}
\includegraphics[width=6cm]{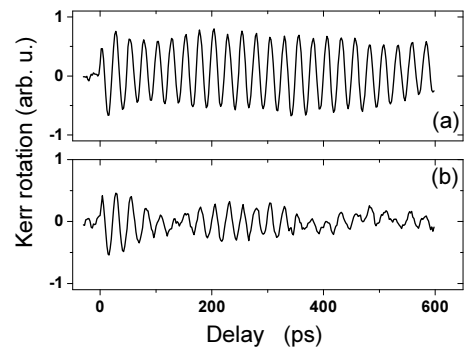}
\end{center}
\caption{TRKR signals obtained with a 1.6~${\rm \mu m}$
wide strip of 2DEG sample, obtained in a modified
experimental setup where non-overlapping pump and probe
spots of $\sim 2 \;  {\rm \mu m} $ diameter were used. The
trace in (a) is obtained with the pump focussed on the wire
and the probe next to the wire with 4~$\mu$ pump-probe
separation. The trace in (b) is obtained with both the pump
and probe focussed on the wire, again with 4~${\rm \mu m}$
pump-probe separation. Data obtained at 4.2~K and 7~Tesla,
pump intensity of $\sim 10 \cdot 10^{11} \; {\rm
photons/cm^2} $ per pulse.} \label{fig:wiremicroscope}
\end{figure}

Secondly, we took TRKR data in an experiment where a
microscope function\cite{rizo2008rsi} was included in the
setup with Laser 1. In these experiments we measured with
pump and probe spots with a diameter of 1.6~$\mu$m that
could be focussed at different (non-overlapping) locations
on the sample. We measured on heterostructure material that
had been processed into an ensemble of parallel wires of
1.2~$\mu$m width at a periodicity of 1.6~$\mu$m. The wires
were realized with electron-beam lithography and subsequent
wet etching to a depth ($\sim 100$~nm) that removes doping
layer (and thereby the 2DEG) between wires.
Figure~\ref{fig:wiremicroscope} shows results that were
obtained with this Kerr microscope. TRKR signals that were
obtained with the pump and probe spot separated by 4~$\mu$m
distance but focussed on the same wire show both population
A and C (beatings in Fig.~\ref{fig:wiremicroscope}b, and
Fourier analysis gives g-factors 0.34 and 0.43). However,
when repeating the measurement with the pump and probe
separated by the same distance, but now separated along a
direction orthogonal to the wires, the TRKR signals only
show population A (no beatings in
Fig.~\ref{fig:wiremicroscope}a, and Fourier analysis gives
only $|g| \approx 0.43$). This shows that population C must
be in the accumulation layer, and that the signal
contribution with $|g| \approx 0.43$ is (at least in the
latter experiment) dominated by electrons in the substrate,
just below the multilayer buffer. When pumping on a wire,
photo-electrons in the accumulation later (II) cannot
escape from the wire: Escape into a deeper layer is blocked
by the multilayer buffer, while lateral escape is blocked
by a $\sim0.5$~eV barrier that results from upward bending
of the conduction band below the etched tranches.
Consequently, no population can appear in the accumulation
layer (II) of neighboring wires. Photo-electrons below the
multilayer buffer, however, can diffuse to areas below
neighboring wires. Note that we cannot conclude from this
experiment that population A is in the substrate (layer I)
in our experiments with the larger and overlapping pump and
probe spots as well. With the microscope we probe only
those electrons that have migrated away from the pump area,
and we pumped at a slightly higher intensity.

In the experiments with large overlapping spots, the clean
segregation into two distinct signal components
nevertheless points to the interpretation that population A
resides in the substrate (and population C in the 2DEG).
This interpretation then assumes that all photo-electrons
that are induced throughout the accumulation layer (II)
rapidly drift into heterojunction quantum well (see
Fig.~\ref{fig:schematic}b), and the segregation between
signal components A and C then results from the fact that
the multilayer buffer blocks this drift for photo-electrons
in the substrate. However, Salis \textit{et al.} recently
reported that the mobility of photo-electrons along the
growth direction is surprisingly low at low pump-photon
densities \cite{salis2006prl}. This means that
photo-electrons near the multilayer buffer in the
accumulation layer (II) do not migrate into the
heterojunction quantum well within a few picoseconds.
Instead, these electrons then have a dwell-time for staying
in the deeper half of layer II that is in excess of the
observed Kerr signal decay times, and give a contribution
to Kerr signals with $|g|\approx 0.44$. This scenario is
also consistent with the fact that the amplitude of signal
contribution A is comparable to that of C (unlike the
results for the strong-pumping regime, where B gives a
$\sim$5 times stronger contribution than A, see the
discussion in Section~\ref{subsec:GaAsKerr}). When
population A is in the substrate (I), its signal is
expected to be much weaker than the signal from C due to
absorption throughout the accumulation layer and multiple
reflections on the multilayer buffer (at least in the
mono-color experiment, discussed below). This analysis
point to the conclusion that at low pump-photon densities
signal contribution A is dominated by photo-electrons just
above the multilayer buffer (in layer II). However, we
cannot fully rule a contribution from electrons below the
multilayer buffer (in layer I), since the low mobility
values are not yet fully understood.

For both the experiment with large laser spots and the
microscope, we could again only observe two populations
when pumping well above the bottom of the conduction band
for bulk GaAs. With the pump wavelength longer than 808~nm,
the Kerr signals were dominated by a single population with
$|g| \approx 0.44$. The population with $|g| \approx 0.34$
could be observed in both the two-color (Laser~1) and
monocolor (Laser~2) experiment when pumping with 808~nm or
shorter pump wavelength (see also the discussion in
Section~\ref{subsec:iGaAsKerr}). However, different than
for the experiments at high pump-photon densities, it is
now not only a matter of having a short penetration depth
for the pump. Obtaining Kerr signals with a strong
contribution from 2DEG electrons probably also requires
direct pumping into 2DEG subbands, and such transitions
only obtain a significant matrix element when pumping at
least $\sim$20~meV above the bottom of the conduction band
for bulk GaAs. This clearly must play a role when the drift
of photo-electrons into the heterojunction quantum well is
indeed a surprisingly slow process at low pump intensities
\cite{salis2006prl}.

Also for this low pump-photon density regime we studied how
the Kerr signals depend on temperature $T$ (presented in
Fig.~\ref{fig:fitparamvstemp}) and magnetic field $B$ (only
discussed). The Kerr signal decay times for populations A
and C as a function of temperature are presented in
Fig.~\ref{fig:fitparamvstemp}c, and the corresponding
g-factors in Fig.~\ref{fig:fitparamvstemp}d, as obtained
from fitting Eq.~\ref{eq:fitfunction} on TRKR measurements
taken at $\sim3.6\cdot 10^{11} \: {\rm photons/cm^2}$ per
pump pulse. The temperature dependence of the g-factor of
population A follows again the trend that was reported for
bulk GaAs \cite{oestreich1996prb}, and agrees with the high
pump-photon density results when accounting for the lower
photo-electron density. The trend for C shows a weaker
temperature dependence, in agreement with a weaker
dependence on $E$ for 2DEG electrons
(Eq.~\ref{eq:gfactorvsE}).

For the Kerr signal decay times
(Fig.~\ref{fig:fitparamvstemp}c), we found evidence that at
4.2 K the decay rate for both population A and C is
dominated by the D'Yakonov-Perel' (DP) spin dephasing
mechanism. We conclude this from further experiments where
our 2DEG material was etched into 1.2~$\mu$m wide wires.
The decay times then show a clear dependence on the crystal
orientation of the wire. This proves that the DP mechanism
dominates, with spin-orbit fields that are highly
anisotropic in $k$-space. This results from a Rashba
contribution \cite{zutic2004rmp} to the spin-orbit fields
that can cancel the Dresselhaus contribution. The details
of this study will appear in a future publication
\cite{denega2009arxiv}. For the 2DEG population this
conclusion agrees with earlier studies on high-mobility
2DEGs \cite{brand2002prl,miller2003prl,stich2007prl}. The
fact that the decay time for C shows almost no dependence
on temperature, indicates that the DP mechanism remains
dominant up to 100 K, and that in this low pump-intensity
regime the 2DEG mobility is not limited by scattering on
acoustic phonons for all these temperatures. The Kerr
signal decay times for population A are around 40~K longer
than the electron-hole recombination times for bulk (as
measured on bulk i-GaAs). This provides another indication
that population A resides in a layer around the multilayer
buffer where electron-hole recombination is suppressed due
to band bending. The increase of the decay time for A when
increasing the temperature up to 40~K indicates a
suppression of the DP mechanism due to more rapid momentum
scattering. Apparently, momentum scattering for this
population is due to acoustic phonons. In the range 40~K to
100~K, the decay times shorten again. Here, the thermal
spread in electron kinetic energies $E$ becomes important,
and results in a spread in g-factor values $\Delta g$ that
increases the precessional dephasing rate.

\section{\label{sec:conclusion}Summary and Conclusions}

We have characterized under what conditions it is possible
to use time-resolved Kerr rotation measurements for studies
of electron spin dynamics of 2DEG and epilayer ensembles in
a GaAs/AlGaAs heterojunction system with a high-mobility
2DEG. Differences in g-factors of electrons in different
layers of the heterostructure allow for discriminating the
various populations, and for defining regimes where they
behave as independent populations. This technique can be
applied thanks to rapid momentum relaxation processes,
which allows for pumping photo-electrons with an excess
energy into the conduction band. At high pump-photon
densities (above the 2DEG density from doping), this is
crucial for having pump pulses with a short penetration
depth, which ensures that Kerr detection is mainly probing
electron populations near the wafer surface. At low
pump-densities, where 2DEG spins can be observed, the
excess pump-photon energy is also needed for directly
pumping into subbands of the 2DEG. Furthermore, we find
that having a barrier at about 1~$\mu$m depth in the
accumulation layer facilitates the segregation between Kerr
signals. At high pump-photon densities this directly blocks
the intermixing of accumulation-layer and substrate
populations, and this drives the difference in g-factor.
Here the accumulation layer can be studied as a bulk
epilayer filled with electrons. Further, since the Kerr
response is enhanced by reflection on each interface in the
heterostructure, the multilayer buffer may also be
important at low pump-photon densities for realizing a
segregation between the Kerr response from electrons in the
heterojunction quantum well and electrons around the
multilayer buffer.

We analyzed that in our experiments several mechanism
contribute to the dephasing of spin-oriented electron
populations and the decay of Kerr signals, and complete
understanding requires further studies. Our results here
provide a basis for such further investigations, in
particular for the spin dynamics of 2D electron ensembles
in heterojunction quantum wells. An interesting direction
is to explore whether confinement of such electron
ensembles in micronscale device structures (as large
quantum dots and wires) can be used to study spin dephasing
anisotropy and suppression of spin dephasing due to
confinement \cite{liu2009jsnm,denega2009arxiv}. In
addition, continuing our work with a microscope function in
the setup allows for studying spin dynamics of 2DEG spins
in transport channels with very high time resolution and
micronscale spatial resolution \cite{rizo2008rsi}. Our
results thereby provide a path to performing studies on
heterojunction 2DEG channels with tunable spin-orbit
effects and electron density, which is interesting for work
on the Datta-Das spin-transistor concept
\cite{datta1990apl}. Notably, we showed that this can also
be applied with non-overlapping pump and probe, giving
access to exploring correlations between spin dynamics and
spin transport.

\begin{acknowledgments}
We thank Bernard Wolfs, Ji Liu, and Thorsten Last for help
and useful discussions, and Bernd Beschoten for providing
the bulk $n$-GaAs sample. This work was financially
supported by the Dutch Foundation for Fundamental Research
on Matter (FOM), the Netherlands Organization for
Scientific Research (NWO), and the German programs DFG-SFB
491, DFG-SPP 1285 and BMBF nanoQUIT.

\end{acknowledgments}


\begin{thebibliography}{99}
\bibitem{gfactordriftvalue} For the $n$-GaAs
sample and bulk populations in the heterostructure sample,
we observe g-factors close to $|g| = 0.44$. At the highest
pump intensities, however, we observes values up to
$|g|\approx 0.42$, as can be expected from band filling
(see. Eq.~\ref{eq:gfactorvsE}). In addition, we observed
for such high-pump-intensity traces that this effect
diminishes over time scales of the electron-hole
recombination times, resulting in weak g-factor drift
during a TRKR trace that brings the g-factor back towards
$|g| \approx 0.44$. We did not include this in our fitting
as with Eq.~\ref{eq:fitfunction}, since it would give a
much higher number of free parameters. However, this drift
in-g-factor values only gave small shifts ($< \negmedspace
\pi$) in the phase of Kerr signals. We checked that the
extracted $\tau_K$ values with our approach were not
sigificantly different from fitting results where these
effects were included.
\bibitem{phases} We observe that using phases $\phi_i$ as free
parameters is needed for obtaining good fits. We also
observed this with the $n$-GaAs reference material (but not
when pumping at low intensities close to the bottom of the
conduction band). This contradicts with the expected
direction of spin orientation parallel to the pump beam at
$t=0$. However, we only use data from delays $t>5$~ps for
fitting, since holes contribute to the Kerr signals up to
$t \approx 5$~ps. For $t<5$~ps there is also rapid
relaxation of electron kinetic energy $E$. Since the
effective g-factor has a value closer to zero for higher
$E$, early energy relaxation is consistent with apparent
phase offsets for $t>5$~ps. However, full justification of
using $\phi_i$ as free parameters requires further
investigation.
\end{thebibliography}
\end{document}